\documentclass[12pt,epsf,epsfig,psfig]{article}
\usepackage{graphicx}
\usepackage{epsfig}
\usepackage{cite}
\def\e{\epsilon}

\def\be{\begin{equation}}
\def\ee{\end{equation}}

\def\lsim{\raise0.3ex\hbox{$<$\kern-0.75em\raise-1.1ex\hbox{$\sim$}}}
\def\gsim{\raise0.3ex\hbox{$>$\kern-0.75em\raise-1.1ex\hbox{$\sim$}}}

\def\NP{{ Nucl.\ Phys.\ }}
\def\PL{{ Phys.\ Lett.\ }}
\def\PR{{ Phys.\ Rev.\ }}

\def\PRL{{ Phys.\ Rev.\ Lett.\ }}

\def\ZP{{ Z.\ Phys.\ }}
\def\EP{{ Europ.\ Phys.\ J.\ C}}

\begin{document}


~

\bigskip

\centerline{\Large \bf Self-Organized Criticality:}

\bigskip

\centerline{\Large \bf A New Approach to Multihadron 
Production\footnote{Invited Talk given at the 40${\rm th}$ Max Born Symposium,
{\sl Strong Correlations in Dense Matter}, Wroc\l aw/Poland, 
October 9 - 11, 2019.}}

\vskip1cm

\centerline{\large \bf Helmut Satz}

\bigskip

\centerline{\large Fakult\"at f\"ur Physik, Universit\"at Bielefeld, Germany}

\vskip1.6cm

\centerline{\large \bf Abstract}

\medskip

We apply the concept of self-organized criticality in statitistical physics to
the study of multihadron production in high energy physics.

\bigskip

{\large
\section{Introduction}}

In the last half of the past century, the central issues of elementary
particle physics were

\medskip

\begin{itemize}
{\item what are the ultimate consituents of matter?}
{\item what are the forces between them?}
{\item what is the ultimate theory: QCD, electroweak, standard model, 
GUT, TOE?}
\end{itemize}
They reflect the age-old reductionist approach to the study of matter.

\medskip

The transition to the new millenium witnessed, in various degrees of clarity,
a change of paradigm. It can perhaps be best summarized in the words of Per 
Bak 
\cite{Bak}

\medskip

\centerline{\sl The laws of physics are simple, but nature is complex.}

\medskip

Bak went on to ask

\medskip

{\sl How can the universe start with a few types of elementary particles 
at the big bang, and end up with life, history, economics and literature? 
Why did the big bang not form a simple gas of particles or condense into 
one big crystal?}

\medskip

In other words, the new issue was to understand how the structured complexity
of the world around us could arise. Thus, the new concepts determining much of
the work of the past twenty years are
\begin{itemize}
\item{emergence, complexity, fractality, chaos}
\item{non-equilibrium behavior, self-organization}
\end{itemize}
In physics, this has led to more intensive studies of emergent phenomena in
non-equilibrium processes, in mathematics to that of fractal structures.
It has moreover led to a general framework applicable as well to swarm
formation in biology and to financial fluctuations in market patterns.
In this talk, I want to show how it can provide a new view of multihadron
production in high energy collisions \cite{CS}.

{\large
\section{Criticality}}

Let us begin by recalling the standard study of correlations in a system of 
identical constituents (spins, particles, birds,...) subject to next neighbor
interaction. For a given value of the control parameter $T$ (``temperature'')
the correlation of two constituents at a separation $r$ is given by the
correlation function
\be
\Gamma(r,T) \sim {a \over r^p} \exp{-r/\lambda},
\ee
in terms of a dimensional constant $a$, 
an emergent correlation length $\lambda(T)$ and a 
power-law exponent $p \simeq 1$. The correlation is thus scale-dependent,
it becomes weaker with increasing separation,
\be
{\Gamma(2r,T) \over \Gamma(r,T)} = (1/2)^p \exp{-r/\lambda},
\ee 
measured in units of the correlation length. At the critical point of
the system, the correlation length diverges $\lambda \to \infty$, so that
\be
\Gamma(r,T_c) \sim {a \over r^p}
\ee
and hence the relative correlation for two different separations 
becomes scale-independent,
\be
{\Gamma(2r,T_c) \over \Gamma(r,T_c)} = (1/2)^p. 
\ee
The correlation is now independent of the separation $r$, there is no longer
any self-organized scale $\lambda$.

\bigskip

{\large \section{Self-Organized Criticality}}

For systems in equilibrium, we have a control parameter $T$ and an order
parameter $m(T)$. To achieve criticality, an outside operator tunes the
temperature adiabatically, $T \to T_c$, and at $T_c$, $m(T)$ changes abruptly.
In other words, tuning of the control parameter changes the order parameter.

\medskip

Non-equilibrium systems, on the other hand, evolve on their own, there is no
tunig operator. As a result, the order parameter changes.
Given suitable dynamics the evolution drives the system to a
critical point (``critical attractor''). We now have an evolving order
parameter, which results in a changing control parameter.  

\medskip

Per Bak has illustrated this in his now well-known sand-pile scenario
\cite{Bak-sand}. Pouring sand slowly onto a flat surface leads to the
formation of a sand pile, whose slope $G$ continues to increase. When it
reaches a critical value $G_c$, avalanches descend, and as more sand is added,
more avalanches occur. We now record in the course of time the number of
avalanches of size $s$, with the size determined by the amount of sand moved.
The result is

\be
n(s) = \left({a \over s}\right)^p~ \to~ 
\log n(s) = p \log s + {\rm const.};~~~
{n(s)\over n(2s)} = (1/2)^p \not= f(s)
\ee

We thus have a constant input, the pouring of sand, driving the slope to
criticality; the output then concists of avalanches with a power-law size
distribution, leading to scale-invariant size ratios.

\medskip


Another application is given by the study of earthquakes. Measurements in
New Mexico give for the number $n(s)$ of earthquakes of size $s$ on the
Richter scale the form (see ref.\ \cite{Bak}),
\be
\log n(s) = p \log s + {\rm const.}
\ee
Here the input is the increasing pressure in the crust of the earth, the
resulting output the number of earthquakes of size $s$, in the form of a 
power-law in $s$. 

\begin{figure}[h]
\centerline{\psfig{file=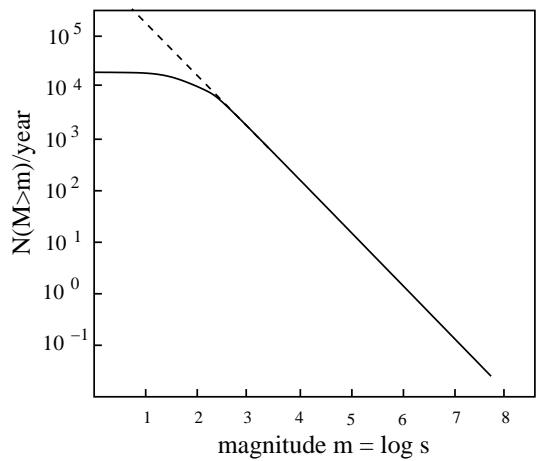,width=7cm}}
\caption{Earthquake distribution in New Mexico}
\end{figure}

The power-law form is seen to be quite well satisfied over six orders of
magnitude. The deviations at low $s$ arise from difficulties in detecting very
weak earthquakes.

\medskip

{\large \section{Scale-Invariance}}

A simple example of scale-invariance is given by the ordered partitioning of
an integer $n$ into integers \cite{blan}, such as
\be
n= 3: 3, 2+1, 1+2, 1+1+1 ~~~~~~~~ q(3)=4;
\ee
for $n=3$, there are $q(n=3)=4$ partitions.
It is readily seen that the general result is
\be
q(n) = 2^{n-1} = {1\over 2} \exp\{n \ln 2\}.
\ee
Note that the unordered case is more difficult, see Hardy and Ramanujan.
We now ask how often the number $k$ occurs in the set of all
partitionings of $n$; we then want to define this number $N(k,n)$ in
terms of a strength $s(k)$ of the number $k$. It seems natural to define 
this strength as the number of partionings of $k$ itself,
\be
s(k) = q(k) = {1\over 2} \exp(k \ln 2).
\ee
In a framework of self-organized criticality, we then expect the number
$N(s(k),n)$ of partionings of strength $s(k)$ to show power-law
behavior,
\be
N(s(k),n) = \alpha(n) [s(k)]^{-p}
\ee
For the partion entropy $S(k,n)= \log N(s(k),n)$ we then obtain the form
\be
S(k,n) =
\log N(k,n)
= -k (p \log e \ln 2) + 
{\rm  const.}(n)~
\ee
In Fig. \label{3}, this is already seen to be well-satisfied for quite small
values of $n$.

\begin{figure}[htb]
\centerline{\psfig{file=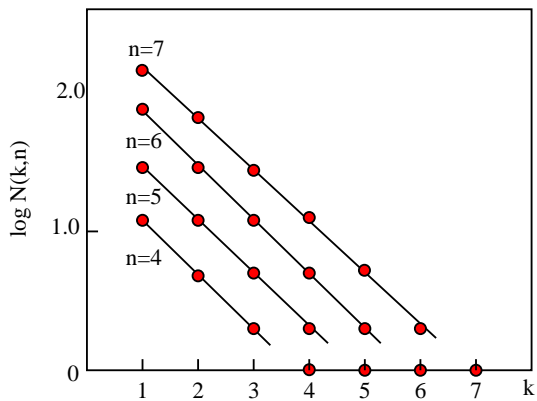,width=7cm}} 
\caption{Partition entropy $S(k,n)$ vs. $k$ for different $n$.}
\end{figure}

\medskip

An integer consists of integers, which consist of integers,...:that brings to
mind Rolf Hagedorn, who postulated that ``fireballs consist of fireballs,
which consist of fireballs,...''. Here we have once more a partitioning 
problem, but a generalized one, in which fireballs consist of moving fireballs,
having kinetic energy, which has to be taken into account in the partitioning.

\medskip

The equation governing the composition of fireballs is Hagedorn's statistical
bootstrap equation \cite{yellow}, stating that there are $\rho(m)$ states of
mass $m$
\be
\rho(m) = \delta(m\!-\!m_0) ~+
\sum_N {1\over N!} \left[ {4 \pi \over 3(2\pi m_0)^3} \right]^{N-1} 
\hskip-0.2cm \int \prod_{i=1}^N ~[dm_i~ \rho(m_i)~ d^3p_i] 
~\delta^4(\Sigma_i p_i - p),
\ee
with $m_0$ for the ground state hadron. Its solution is given by \cite{Nahm}
\be
\rho(m) \sim [1+ (m/\mu_0)]^{-a} \e^{m/T_H} 
~\to~ \ln \rho \sim {m\over T_H}
- a \ln [1 + (m/\mu_0)], 
\ee
with a normalization constant $\mu_0$. The crucial constant determining the
exponential growth, the so-called Hagedorn ``temperature'' $T_H$, is the  
solution of the equation
\be
\left({2\over 3 \pi}\right)\left(T_H \over m_0\right) 
K_2(m_0/T_H) = 2 \ln 2 - 1.
\ee
Note that this Hagedorn ``temperature'' is totally of combinatoric origin;
it corresponds to the ``temperature'' $\Theta =1/ \ln 2$ in the partitioning
of integers. It becomes a temperature only in the partition function of a
resoance gas of states with spectral weights $\rho(m)$.

\medskip

{\large \section{High Energy Hadron Production}}

The aim now is to formulate hadron production in high energy collisions as
self-organized criticality \cite{CS}. The initial state just after the 
collision is a beam of energetic colored partons flying along the collision
axis. In their passage, they lose energy by doing work against the physical
vauum: the collision is a non-equilibrium process. The passage sorts the
partons: at small rapidity, we have the slowest, and with increasing rapidity
the faster and faster ones. This corresponds to the well-known
``inside-outside'' cascade of Bjorken \cite{bjorken}.

\medskip

In the conventional scenario describing such collisions \cite{beca,resgas}, 
one considers a slice
at fixed longitudinal rapidity, corresponding to a partial system at fixed
time. It is assumed that the resulting system is a bubble of deconfined medium
(QGP) in full local equilibrium. This bubble expands, cools, and then
eventually hadronizes. At this point, one assumes chemical freeze-out, the end
of chemical equilibrium. The resulting interacting hadronic medium is assumed
to still be in local thermal equilibrium. It continues to expand until it
reaches kinetic freeze-out, leading to free hadrons. This is the end of
thermal equilibrium. 

\medskip 

The hadronization transition of the QGP is studied in finite temperature
lattice QCD \cite{baza1}.
Close to the chiral limit of vanishing quark masses ($m_q \to
0$), one finds critical behavior, with a correlation length $\lambda(T,m_q=0)$
diverging at the critical point $T=T_c$,
\be
\lambda(T,m_q=0) \sim |T-T_c|^{-\nu}.
\ee
As a result, the correlation function become scale-invariant,
\be
 \Gamma(r,T_c)/\Gamma(2r,T_c) = 2^p, 
\ee
i.e., independent of the separation distance $r$.

\medskip

For small finite quark masses, a reflection of criticality remains, one has
``pseudo-critical'' behavior at a temperature $T_c\simeq 155$ MeV

\begin{figure}[htb]
\centerline{\psfig{file=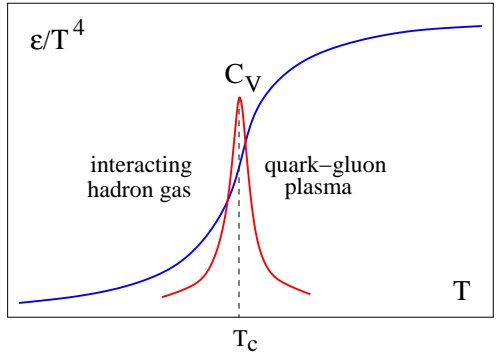,width=6.5cm}} 
\caption{Pseudo-critical behavior for the energy density $\epsilon$
and the specific heat $C_v$.}
\end{figure}

\medskip

If the local bubbles are adiabatically evolving QCD matter, i.e., if the
evolution occurs in equilibrium, with
a QGP above and 
an interacting hadron gas below $T_c$, then the relative hadron abundances
in the hadronic medium below $T_c$ are not fixed; they decrease as

\be{\phi(m_i,T)\over \phi(m_j,T)} 
\sim \exp-\left({(m_i - m_j)\over T}\right).
\ee

In other words, the adiabatic evolution of a QGP does not lead to chemical
freeze-out at $T_c$. Moreover, the hadron masses are also still temperature-
dependent, as given by 
\be
{m_h(T)\over m_h(0)} \sim {f_{\pi}(T) \over f_{\pi}(0)},
\ee
where $f_{\pi}(T)$ corresponds to the pion decay constant.  

\medskip

These features of an interacting QCD medium in equilibrium are found to
disagree with high energy hadroproduction data. One there observes that the
relative abundances of the different hadron species are fixed by the yields
at $T_c$, specified by the corresponding Boltzmann factors $\phi(m_i,T_c)$,
with vacuum masses $m_i$ for the species $i$. In other words, an ideal gas of
all hadronic resonances with vacuum masses at $T_c$ correctly predicts ``all'' 
abundances \cite{resgas}. The ``all'' implies some caveats.

\medskip

Hadrons containing heavy flavors (charm and bottom) cannot be directly 
compared to those made up of light flavors, since their perturbative hard
production process results in a different energy dependence than that of
light flavor hadrons. In elementary collisions ($e^+e^-,~pp$), the ideal
resonance gas otherwise accounts for all hadrons, In $AA$ collisions, it
does so for stable hadrons, while resonance ($\rho,~K^*, N^*$) can still
suffer modifications due to the overlap of nucleon-nucleon interactions.

\medskip

The conventional scenario thus encounters two immediate basic difficulties. 

1) Why is there chemical freeze-out directly at $T_c$? 

2) Why are the abundance ratios at $T_c$ those based on vacuum masses? 

\medskip

A recent further and perhaps indicative problem was triggered by recent
$Pb-Pb$ LHC data by the ALICE collaboration. They find that even the
abundances of light nuclei (deuteron, triton, helium) are correctily predicted
by the ideal Boltzmannn gas at $T_c$. 

\medskip

These states are both large
(a triton has a size of the total interaction region) and very loosely bound,
so that:

3) They cannot exist in an interacting hadronic medium of temperature
$T_c$.  

We believe that with these three points nature is trying to tell us that a
scenario, in which hadronization produces an interacting hadronic medium
of temperature $T \sim T_c$ and with vacuum masses, cannot be correct, and so
we look for an alternative.

\bigskip

{\large\section{The SOC Scenario}}

We thus consider a non-equilibrium parton system, the counterpart of pouring
sand, converging towards a pseudo-critical point, at which it breaks up into
all permissible hadron states - the avalanches. More specifically, in the 
absorptive state form of SOC, the {\sl colored} parton state undergoes {\sl 
color absorption} at the pseudo-critical point, giving rise to all possible 
{\sl color neutral} hadron states. 
   
\medskip

The crucial difference to the conventional scenario of a hadronizing QGP is
that in SOC the hot colored partonic medium is quenched by the cold    
color-neutral vacuum, breaking up into free color-neutral hadrons, without any
subsequent hot interacting hadronic medium.

\medskip 

In SOC, the number $N(m)$ of produced hadrons of mass $m$, is described by the
scale-invariant form

\be
N(m) = \alpha [\rho(m)]^{-p}
\ee
 
in terms of the resonance strength $\rho(m)$. We assume that $\rho(m)$ is
given by the composition law of states in the Hagedorn bootstrap,

\be
\log N(m) = -m \left({p \log e \over T_H} \right) \left[1 
- \left(a T_H \over
m \right ) 
{\ln(1+{m \over \mu_0})}\right]
+ {\rm const.} 
\ee

Let us compare the ALICE data for Pb-Pb collisions at $\sqrt s = 2.76$ GeV
\cite{alice}
to a somewhat simplified form

\be
\log[(dN/dy)/(2s+1)] \simeq -m\left({\log e~\!p\over T_H}\right) + A,
\ee

with $T_H=155$ MeV and fit values $p=0.9$, $A=3.4$

\begin{figure}[htb]
\centerline{\psfig{file=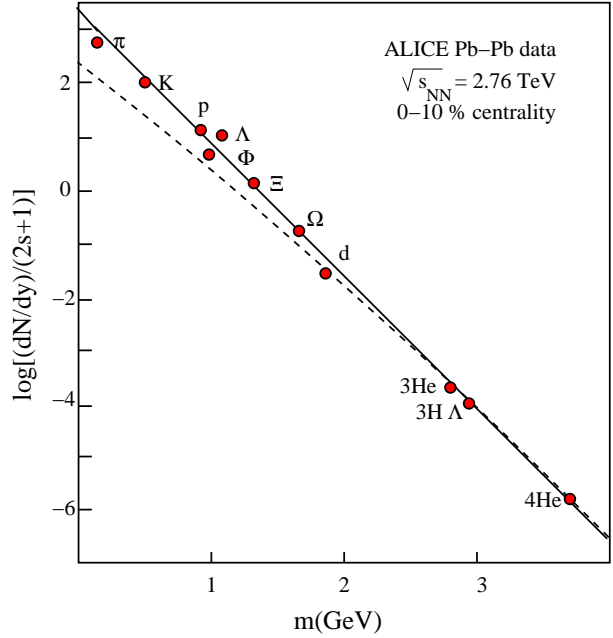,width=8cm}} 
\caption{ALICE data \cite{alice}
compared to SOC predictions; full line eq.\ (21), dashed
  line eq. (20).}

\end{figure}
 
Including the missing correcton terms leads to the dashed line; the difference
between the two curves accounts for the production elementary hadrons through
resonance decay.

\medskip

For elementary collisions, such as $p-p$, we expect the avalanches to consist
of different individual hadrons. In high energy $A-A$ collisions, overlap and
interactions of the ``debris'' can in fact affect resonance production,
leading to further modifications of rates for $\rho,~K^*,~\Delta$ etc. The
results of this are not the finite temperature equilibrium hadron gas of
QCD. 

\bigskip

{\large\section{Conclusions}}

We have proposed that in high energy collisions, a non-equilibrium colored
parton beam converges as a function of rapidity towards a pseudo-critical
attractor, the color absorbing state. At that point, quenching leads to color
neutrality in form of an avalanche of hadrons, with scale-invariant mass
distributions. With evolving rapidity, there are successive avalanches, and
the sum over all hadron distributions produces a thermal distribution at
the pseudo-critical temperature $T_c$.

\bigskip

{\large\section{Acknowledgement}}

The material presented here was developed together with Paolo 
Castorina \cite{CS},
and I am very grateful to him for numerous stimulating
discussions.

\end{document}